\documentclass[useAMS,usenatbib]{mn2e}    

\def\msun{\mbox{M$_{\odot}$}}
\def\msunpc{\mbox{M$_{\odot}$pc$^{-3}$}}
\def\fdunit{\mbox{M$_{\odot}$(pc km/s)$^{-3}$}}
\def\fd{\mbox{$f'_{\rm d,inn}$}}
\def\fb{\mbox{$f'_{\rm b}$}}
\def\Rdi{\mbox{R$_{\rm d,in}$}}
\def\fp{\mbox{$f^{\prime}$}}

\def\kms{\mbox{km s$^{-1}$}}

\def\lesssim{{_ <\atop{^\sim}}}
 
\title[Secular evolution of galactic discs]{Secular evolution of galactic discs: 
constraints on phase-space density} 
\author[Avila-Reese et al.]{V. Avila-Reese$^{1}$, A. Carrillo$^{1}$, 
O. Valenzuela$^{2}$ and A. Klypin$^{3}$ \\
$^{1}$Instituto de Astronom\'\i a, Universidad Nacional Aut\'onoma de M\'exico, A.P. 70-264, 04510 M\'exico, D.F. \\
$^{2}$Department of Astronomy, University of Washington, Box 351580, Seattle, WA 98195 \\
$^{3}$Astronomy Department, New Mexico State University, Las Cruces, NM 88001}

\begin{document}


\pagerange{\pageref{firstpage}--\pageref{lastpage}} \pubyear{2004}

\maketitle

\label{firstpage} 

\begin{abstract}
It was argued in the past that bulges of galaxies cannot be formed 
through collisionless secular evolution because that would violate
constraints on the phase-space density: the phase-space density in bulges 
is several times larger than in the inner parts of discs. We show that 
these arguments against secular evolution are not correct. Observations 
give estimates of the coarsely grained phase-space densities of galaxies,
\fp$=\rho_s/\sigma_R\sigma_{\phi}\sigma_z$, where $\rho_s$ is stellar
density and $\sigma_R, \sigma_{\phi}, \sigma_z$ are the radial,
tangential, and vertical rms velocities of stars. Using high-resolution 
N-body simulations, we study the evolution of \fp\ in stellar discs of 
Galaxy-size models. During the secular evolution, the discs,
which are embedded in live Cold Dark Matter haloes, form a bar and
then a thick, dynamically hot, central mass concentration. In the
course of evolution \fp~ declines at all radii. However, the decline 
is different in different parts of the disc. In the inner disc, \fp$(R)$ 
develops a valley with a minimum around the end of the central mass 
concentration. The final result is that the values of \fp~ in the central 
regions are significantly larger than those in the inner disc. The minimum, 
which gets deeper with time, seems to be due to a large phase mixing produced 
by the outer bar. We find that the shape and the amplitude of \fp$(R)$ for 
different simulations agree qualitatively with the observed \fp$(R)$ in 
our Galaxy.  Curiously enough, the fact that the coarsely grained phase-space 
density of the bulge is significantly larger than the one of the inner
disc turns out to be an argument in favor of secular formation of bulges, 
not against it.
\end{abstract}


\section{Introduction}

Formation of galactic spheroids remains as a major unsolved problem in
astronomy. This is an important problem, specially if one takes
into account that at least half of the stars in the local Universe are 
in spheroids: either bulges or ellipticals (e.g., Fukugita, Hoggan \& 
Peebles 1998; Bell et al. 2003). The key question is how and where these 
stars formed. 
One possibility is that stars in present-day spheroids were formed in a 
self-regulated quiescent fashion characteristic of 
galactic discs, and then the disc stars were dynamically
heated by mergers and/or secular disc processes. In this case
the spheroid formation is predominantly collisionless.  Another
possibility is that spheroid star formation (SF) was highly dissipative
and proceeded in a violent, possibly bursting and dust enshrouded mode
during a dissipative disc merging event or during a phase of fast 
gas (monolithic) collapse. Although both possibilities happen certainly
in the real Universe, it is important to evaluate the feasibility
of each one as well as the physical/evolutionary context in which one or 
another possibility dominates. In the former case the SF rate is traced by
UV/optical emission, while in the latter by FIR/submillimetre emission. 
Thus, understanding the mechanisms of spheroid formation
and the regimes of formation of their stars is of crucial relevance for 
interpreting and modeling the contribution of present-day stars
in spheroids to the cosmic SF rate history.

\subsection{Secular bulge formation mechanism}

In this paper we will study some aspects of the disc secular evolution.
According to the secular scenario, the formation of a central
mass concentration (a bar or a pseudobulge) happens in a
predominantly dissipationless fashion in the course of development of
gravitational instabilities in the central region of a galactic
stellar disc.  The evolution of the bar can give rise to a central
component that is denser and thicker than the initial thin stellar
disc (Kormendy 1979, 1982).  In earlier simulations the bar in most of
cases was dissolving, leaving behind a pseudobulge (e.g., Combes \&
Sanders 1981; Pfenniger \& Norman 1990; Combes et al. 1990; Raha et
al. 1991; Norman, Sellwood \& Hassan 1996).  However, more recent
simulations, which have many more particles and have more realistic
setup, do not produce typically decaying bars (Debattista \& Sellwood
2000; Athanassoula \& Misiriotis 2002; O'Neill \& Dubinsky 2003;
Valenzuela \& Klypin 2003, hereafter VK03; Shen \& Sellwood 2003;
Debattista et al. 2004). In those simulations bars typically slightly
grow over billions of years.

In the VK03 simulations of discs inside live Cold Dark Matter (CDM)
haloes, the redistribution of the angular momentum of the stellar disc
is driven by the evolving bar and by interactions with the dark matter halo.
 This evolution produces a dense central mass concentration with
nearly exponential profile, which resembles surface brightness profiles
of late-type galaxy bulges (see also Athanassoula \& Misiriotis 2002;
O'Neill \& Dubinsky 2003).  Shen \& Sellwood (2003) and Debattista et
al. (2004) argued that neither a small central mass concentration
(e.g. black hole) nor the buckling instability are efficient enough to
destroy a bar.

Whether the bar is destroyed or not, the heating of the central parts
of the stellar disc and accumulation of mass at the centre are common
features in all models of secular evolution. Further exploration,
including a wide range of realistic initial conditions and inclusion
of processes such as gas infall (e.g., Bournaud \& Combes 2002), minor
mergers and satellites (Aguerri et al. 2001), hydrodynamics, SF and
feedback are certainly necessary.
All the processes are likely to play some role in evolution of
galaxies.  Here we intentionally do not include these complex
processes in order to isolate the effects of the secular evolution: we
include only stellar and dark matter components and do not consider any
external effects.

Models of secular disc evolution gradually find their place in the
theory of galaxy formation. Encouraging results were obtained when a
prescription for secular bulge formation was incorporated in CDM
semi-analytical models of disc galaxy formation and evolution (Avila-Reese 
\& Firmani 2000,1999; see also van den Bosch 1998). These models successfully
reproduce the observed correlations of the bulge-to-disc ratio with
other global properties for late-type galaxies. Secular disc evolution 
should be considered as a complementary path of spheroid formation rather 
than a concurring alternative to the dissipative merging mechanism.

From the observational side, an increasing evidence shows that
structural, kinematic, and chemical properties of the bulges
of late-type galaxies are tightly related with properties of the inner
discs (for reviews and references see Wyse, Gilmore \& Franx 1997;
MacArthur, Courteau \& Holtzmann 2003; Carrollo 2004; Kormendy \&
Kennicutt 2004).  Besides, bars -- signature of secular evolution -- are
observed in a large fraction of spiral galaxies. These pieces of evidence 
strongly favor the secular evolution scenario.

\subsection{Do phase-space constraints pose a difficulty for the secular mechanism?}

According to the Liouville's theorem, in a collisionless system the
phase-space density $f({\bf x,v})$ is preserved along trajectories of
individual stars.  Thus, one expects that a collisionless system
``remembers'' its initial distribution of $f({\bf x,v})$, and this
can be used to test the secular evolution scenario.  
In fact what is ``observed'' is not $f({\bf x,v})$, but a rough estimate 
of the coarsely grained phase-space density 
\begin{equation}
\fp = \frac{\rho_s}{\sigma_R\sigma_{\phi}\sigma_z},
\end{equation}
where, $\rho_s$ is stellar density and $\sigma_R,\sigma_{\phi},\sigma_z$ 
are radial, tangential, and vertical rms velocities of stars.  The 
coarse-grained phase-space density is not preserved. Still, there are 
significant constraints on the evolution of \fp, which are imposed by the 
mixing theorem (Tremaine, Henon, Lynden-Bell 1986). The process that changes
\fp~ is the mixing. Bringing and mixing together two patches of stars with
different fine-grained phase-densities result in \fp, which is lower than 
the maximum phase-space density of the two patches. In other words, the 
mixing results in reducing \fp.  The only way to increase \fp~ is to 
bring in stars with initially large \fp. Indeed, when a bar forms, there is 
a substantial radial infall of mass to the central region. Yet, this does 
not help much because the bar is formed from the central region of the 
disc where the initial \fp~ is low. Additional mixing produced by the bar 
seems to make the things even worse by lowering down already low \fp.

Simple estimates for elliptical and spiral galaxies indicate that the 
coarse-grained phase space-densities in spirals are lower than in
ellipticals (Carlberg 1986), making difficult to produce elliptical galaxies 
by merging of stellar discs (Hernquist, Spergel \& Heyl 1993).
Wyse (1998) discussed similar arguments but for the Galactic disc in the
context of the secular bulge formation scenario. For an exponential
disc with constant height $h_z$ and with an isotropic velocity-dispersion
tensor one finds that $\fp(R)\propto\rho_s/\sigma_z^3\propto$
($\Sigma_s/2$h$_z$)/($\Sigma_s$h$_z$)$^{3/2}\propto \exp(R/2h_d)$,
where $\Sigma_s$ and $h_d$ are the disc central surface brightness and
scale-length radius, respectively. Therefore, \fp$(R)$ is lower toward
the centre. Wyse (1998) then states:  ``one should not find a higher 
phase-space density in stellar progeny, formed by a collisionless process, 
than in its stellar progenitor''. Observational inferences for our Galaxy 
show actually that \fp~ is higher in the bulge than in the inner disc.  
This discrepancy led Wyse (1998) to conclude that the secular scenario has 
a serious difficulty, unless dissipative physics is included.  

The problem is partially mitigated if one considers that the 
Toomre parameter $Q$ is constant along the initial disc. For $Q=$const, 
the rms velocity is $\sigma_R \propto \exp(-R/h_d)/\kappa(R)$, where the 
epicycle frequency $\kappa$ increases as $R$ decreases. In this case 
$\fp(R)$ has an $U-$shaped profile with a maximum at the centre 
(Lake 1989) and a valley in the inner regions. How steep or shallow is 
this valley depends on the inner behavior of $\kappa(R)$. 

The situation is actually more complicated than the simple picture
outlined above because of the spacial and dynamical properties 
of the system evolve. Formation of bars 
is a complex process that affects a large fraction of the disc -- not 
just the central region. Thus, to study  the overall evolution of \fp$(R)$
one needs to turn to numerical simulations.

The main question, which we address in this paper is how the macroscopic
(observational)  
phase-space density profile, \fp, of stellar discs inside CDM haloes evolves 
during the formation (and potential dissolution) of a bar, and whether the
shape of this profile agrees with estimates from observed disc/bar/bulge galaxy 
systems.  We analyze state-of-the-art high-resolution N-body simulations of
Galaxy-like discs embedded in live CDM haloes. The secular evolution of the 
disc in these simulations yields bars that redistribute particles and produce 
a dynamically hot mass central concentration. 

In \S 2 we present a brief description of the simulations and the procedure to 
estimate the phase-space density. The results are given in \S 3, and in \S 4 
we discuss some aspects of the simulations. In \S 5 a comparison with 
observational estimates is presented. Our summarizing  conclusions are given 
in \S 6.

\section[]{MODELS AND SIMULATIONS}  

We study four N-body simulations for evolution of bars in stellar discs embedded 
in live CDM haloes. Two simulations, $A_1$ and $C$, are taken 
from VK03 and the other two, $D_{\rm hs}$ and $D_{\rm cs}$ are from Klypin 
et al. (2005).
The models were chosen to cover some range of initial conditions and 
parameters, with the aim to test the sensitivity of the results to them. 
For example, the initial Toomre parameter $Q$ is constant along 
the disc for models $C$, 
$D_{\rm hs}$, and $D_{\rm cs}$. The $Q$ parameter is variable for model 
$A_1$ (increases in the central regions). Instead, this model is initially
set to have a radial velocity dispersion as $\sigma_R^2(R)\propto \exp(-R/$h$_d)$. 
As the result, all the models have different initial profiles of the azimutally 
averaged coarsely grained phase-space density. Parameters of the models and 
details of simulations are presented in Table 1.

\begin{table}
  \caption{Parameters of models}
  \begin{tabular}{@{}lcccc}
  \hline

   Parameter                     & $C$   &  $A_1$& $D_{\rm hs}$& $D_{\rm cs}$ \\
 \hline
Disc mass ($10^{10}\msun$)       & 4.8   & 4.3   & 4.35   & 5.0 \\
Total mass ($10^{12}\msun$)      & 1.0   & 2.0   & 1.22   & 1.4 \\
Initial disc scale-length (kpc)  & 2.9   & 3.5   & 2.25   & 2.57 \\
Initial Toomre parameter {\em Q} & 1.2   & $<1.2>$   & 1.8   & 1.3 \\
Initial disc scale-height (kpc)  & 0.14  & 0.25  & 0.17  & 0.20 \\
Halo concentration $c_{NFW}$     & 19.0  & 15    & 18    & 17 \\
Number of disc particles ($10^5$)& 12.9    & 2.0   & 4.6   & 2.3 \\
Number of halo particles ($10^6$)& 8.48   & 3.3   & 3.3   & 2.2  \\
Particle mass ($10^{5}\msun$)    & 0.37  & 2.14  & 0.93  & 2.14\\
Formal force resolution (pc)     & 100   & 22    & 19    & 22\\
\hline
\end{tabular}

\medskip The particle mass refers to the mass of ``stellar'' disc particles. 
This is also the mass of the least massive halo particles.  
\end{table}

The initial conditions are generated using the method introduced 
by Hernquist (1993). The galaxy models  initially have exponential discs
in equilibrium inside a dark matter halo with a density profile 
consistent with CDM cosmological simulations (Navarro, Frenk \& White 1997).
The halo concentrations are set somewhat larger than the expected
concentration $c_{NFW}\approx 12$ for a halo without baryons, which should 
host our Galaxy (Klypin et al. 2002). This is done to mimic the adiabatic
compression of the dark matter produced by baryons sinking to the
centre of the halo in the process of formation of the galaxy.

The haloes are sampled with particles of different masses: particle
mass increases with distance. The lightest dark matter particles have
the same mass as the disc particles.  At any time there are very few
large particles in the central 20-30~kpc region.  The time steps of
simulations were forced to be short. For example for model $C$,
the minimum time step is $1.2\ 10^{5}$~yrs, while for model $D_{\rm hs}$
it is $1.5\ 10^{4}$~yrs. For a reference, in model $C$ the typical 
time that a star requires to travel the vertical disc extension of the 
disc at the radius of 8 kpc requires 372 steps, and the orbital period
in the disc plane at the same distance takes 1900 steps.  Model simulations
$C$, $A_1$, $D_{\rm cs}$, and $D_{\rm hs}$ were followed for $\sim
4.4$, 4, 4, and 7 Gyrs, respectively, using the Adaptive Refinement
Tree (ART) code (Kravtsov et al. 1997).

In the models $C$ and $A_1$ the bar is strong even at the end of
simulations without any indication that it is going to die. In the
model $D_{\rm hs}$ the bar is gradually getting weaker, but it is
still clearly visible. The bars typically buckle at some stage producing
a thick and dynamically hot central mass concentration with a peanut 
shape. The model $D_{\rm cs}$ is the only model where bar dissolved
completely and produced a (pseudo)bulge.

The galaxy models used in the simulations are scaled to roughly mimic
the Galaxy. For example, they have realistic disc scale lengths
($\approx 3$~kpc), scale heights ($\approx 200-300$~pc), and they have
nearly flat rotation curves with $V_c\approx 220$~km/s. Yet, we do not
make an effort to reproduce detailed structure of our Galaxy. For example, 
the radius of the bar in model $C$ is $5-5.5$~kpc -- too large as compared 
with real bar which has radius $3-3.5$~kpc. 

 Model $D_{\rm hs}$ presents a shorter bar ($3$~kpc) and it is a better 
model in that respect\footnote{In order to make a better match of the model
with the Milky Way, we rescaled the model: all coordinates and masses
were scaled down by factor 1.15. As any pure gravitational system, it
can be arbitrary scaled using two free independent scaling factors. In
this case we chose mass and distance; time, velocity, surface
density, and so on are scaled accordingly.}, and we describe it
in more detail to show the reliability of our approach.  
With the exception of disc mass (which is somewhat small), model $D_{\rm hs}$
makes a reasonable match for our Galaxy. Its stellar surface density at 
the ``solar'' distance of
8~kpc is $\Sigma_s=54\msun {\rm pc}^{-2}$. For comparison, for our Galaxy
Kuijken \& Gilmore (1989) find $\Sigma_s=48\pm 8\msun {\rm pc}^{-2}$ while
Siebert et al.(2003) find $\Sigma_s=67\msun {\rm pc}^{-2}$. Stellar rms
velocities in radial and vertical directions in the model are 47 \kms
and 17 \kms. Dehnen \& Binney (1998) give for the old thin disc stellar
population of our Galaxy 40 \kms and 20 \kms, respectively. Within the solar 
radius the model has a ratio of dark matter to total mass of $M_{\rm
DM}/M_{\rm tot}=0.6$.  This ratio is significantly lower inside the bar
radius of 3~kpc: $M_{\rm DM}/M_{\rm tot}=0.35$.  The bar pattern speed
is $\Omega_p=54{\rm Gyr}^{-1}$. Bissantz et al. (2003) give
$\Omega_p=60\pm 5{\rm Gyr}^{-1}$ for our Galaxy although their estimate
is also based on a model.

In order to make a more detailed comparison of the mass distribution
in the model $D_{\rm hs}$ with that of our Galaxy, we mimic the position-velocity 
(P-V) diagram for neutral hydrogen and CO in the plane of our Galaxy. 
Observations of Doppler-shifted 21-cm and CO emission along a line-of-sight 
performed at different galactic longitudes $l$ provide the P-V diagram.  
Because the gas is cold, it provides a good probe for mass profile.  There 
are two especially interesting features in the P-V diagram.  Envelops of the
diagram in the first quadrant ($0<l<90$, $V>0$) and in the third
quadrant ($0>l>-90$, $V<0$) are the terminal velocities (Knapp et al. 
1985; Kerr et al. 1986).
Data in the second and forth quadrants are coming for regions outside
of Solar radius (Blitz \& Spergel 1991).  They have information on
the motion in the outer part of our galaxy. For distances larger than
the radius of the bar 3-4~kpc ($|l|<30^o$) the terminal velocities can
be converted into the circular velocity curve (assuming a distance to
the Galactic center). At smaller distances, perturbations produced by
the Galactic bar are large and cannot be taken into account in a
model-independent way. This is why we use the P-V diagram, not the
rotation curve.
 
We mimic the P-V diagram for the $D_{\rm hs}$ model by selecting ``stellar'' 
particles, that are close to the plane $|z|< 300$~pc and have small velocities
relative to their local environment.  A particle velocity must deviate not
more than ~20\kms from the velocity of its background defined by the
nearest 50-70 particles. This gives the rms line-of-sight velocity of
8~\kms, which is compatible with the random velocities of the cold
gas. We place an ``observer'' in the plane of the disc at the distance
8~kpc. Its position was chosen so that the bar major axis is 20
degrees away from the line joining the ``observer'' and the galactic
center. The observer has the same velocity as the local flow of ``stars''
at that distance. We then measure the line-of-sight velocity of each
cold particle and plot the particles in the longitude-velocity
coordinates.

\begin{figure}
\vspace{8cm}
\includegraphics{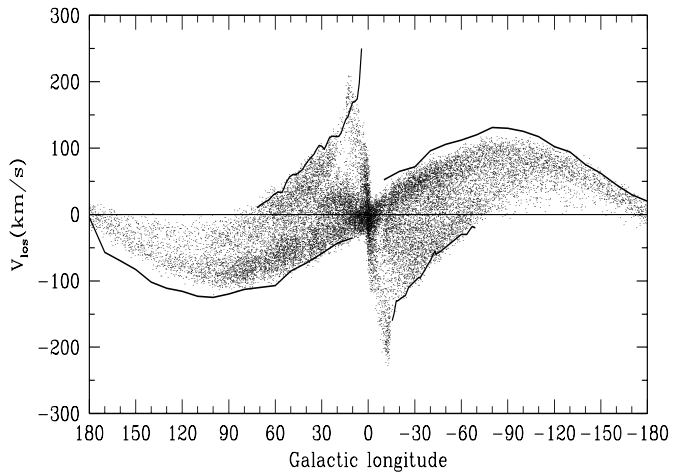}
\caption{Position-velocity (P-V) diagram for the model $D_{\rm hs}$ 
compared with the P-V diagram for neutral hydrogen in our Galaxy. In 
the model the diagram is simulated for an ``observer''
placed at the solar position: in the plane of the disc at the distance
8~kpc from the center.  The full curves in the diagram are the
envelops of the P-V diagram for neutral hydrogen and CO in the Galaxy. 
The envelops in the first (top-left) and third quadrants are
the terminals velocities (Knapp et al. 1985; Kerr et al. 1986),
which define the rotation curve of the Galaxy. The model reproduces
those velocities remarkably well.  The curves in the second and forth quadrants
(Blitz \& Spergel 1991) are related with gas motion in the outer
parts of the Galaxy and with the solar rotation. Deviations of the
curves from the envelop of the points in the model are seen as white
gaps in the second quadrant at large $l$ and in the forth quadrant are
small $l$. The deviations are due to lopsidedness of our Galaxy. }
\label{fig:PV}
\end{figure}

With the procedure just described we are selecting a population 
of ``stellar'' particles, which has small asymmetric drift and
yet has the bulk flows induced by the bar.  The procedure is
insensitive to particular set of parameters as long as rms velocities
stay significantly smaller than the rotational velocity 220\kms and as
long as the radius for the background particles is significantly
smaller than the distance to the center.

Figure~\ref{fig:PV} shows the P-V diagram for the cold ``stellar''
particles in model $D_{\rm hs}$. The envelop of the diagram very 
closely follows that of
our Galaxy, indicating that the mass distribution in the inner part of
the model is compatible with the data on our Galaxy. Small deviations
in the outer part of the galaxy are due to lopsidedness of our Galaxy
(Blitz \& Spergel 1993), which our model cannot reproduce.

\begin{figure}
\vspace{8cm}
\includegraphics{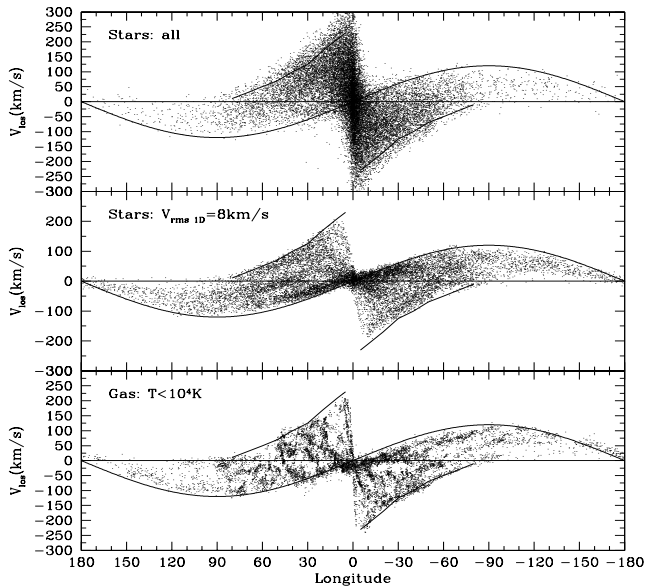}
\caption{Position-velocity diagrams for the N-body+SPH simulation
carried out with the Gasoline code. 
Upper, medium and lower panels are for all the ``stellar'' particles, for only 
cold ``stellar''  particles, and for cold gas, respectively. The full curves 
in every panel are the same. They show the envelops for the cold gas.}
\label{fig:PVtest}
\end{figure}

We assumed that cold ``stellar'' particles resemble neutral and molecular
gas in the P-V diagram.  To test this assumption we use a simulation 
presented elsewhere (Valenzuela et al. 2005, in preparation), which includes 
not only collisionless particles (``stellar'' and dark matter), but also 
gas. This simulation has been run with the Gasoline N-body+SPH code 
(Wadsley, Stadel \& Quinn 2004). 
The galaxy model is similar to the one discussed above.  The simulation had 
$5\times 10^5$ dark matter particles, $2\times 10^5$ disc (``stellar'') 
particles, and $5\times 10^4$ gas particles. The force resolution was 200~pc 
for gas and ``stars'', and 600~pc for dark matter. The simulation was run 
until 1.6~Gyrs.  The simulated ``galaxy'' develops a strong bar with a 
radius $\approx 3~$kpc. We selected an ``observer'' at distance 8~kpc from 
the center at the angle of 20$\circ$ relative to the major axis of the
bar. Figure~\ref{fig:PVtest} shows P-V diagrams for different
components.  P-V diagram for the cold gas with $T<10^4$K shows
remarkable complexity: there are lumps and filaments. Those are due to
spiral arms and shock waves.  The cold ``stellar'' particles do not
have those details, yet they remarkably well follow the same envelops
in the P-V diagram as the cold gas. This is exactly what we want to
demonstrate. The whole stellar population (the top panel) shows large
velocities in the central region -- well in excess of cold gas
motions. In this case, significant (20-30 percent) corrections 
are indeed required to account for the asymmetric drift.

\subsection{Measuring the phase space density}

The  phase-space density is defined as the number of stars 
in a region of phase space around a point ({\bf $x,v$}) divided by the
volume in phase space of that region {\it as} this volume tends
to zero (e.g., Binney \& Tremaine 1987; Wyse 1998). The measurable  
quantity is the coarsely grained phase-space density that is defined in 
finite phase space volumes. However, this quantity is still difficult 
to infer observationally. The measure of the phase-space density commonly 
derived from observations is the stellar spatial density, $\rho_s$,  
divided by the cube of the stellar velocity dispersion (eq. 1). For the 
latter, one typically uses either the projected line of sight velocity 
dispersion, $\sigma_p$, or the inferred radial velocity dispersion, 
 or the product of the three components of the velocity dispersions, 
$\sigma_R, \sigma_{\phi},$ and $\sigma_z$ if they are known.  
We will refer to \fp\ as the azimuthally averaged ``observational'' 
phase-space density as defined above.

Because our main aim is to analyze the coarsely grained phase-space 
density evolution in the N-body simulations in a way that mimics 
observations, we do the following.
We calculate the average density of ``stellar'' particles, $\rho_s$(R), 
and their velocity dispersions within cylindrical (equatorial) rings of 
width $\Delta R$ and thickness $\Delta Z$. Thus, \fp~ is estimated  in a 
representative region above and below the disc plane. Nonetheless, we 
have checked that 
qualitatively similar results are obtained when using other binning or 
even other geometries, in particular the spherical one for the centre. 
The thickness $\Delta Z$ and $\Delta R$ were assumed to be $2h_z$(R) and 200~pc, 
respectively. The results do not change significantly for a large range 
of assumed values for $\Delta Z$ and $\Delta R$. 

\begin{figure}
\vspace{8cm}
\includegraphics{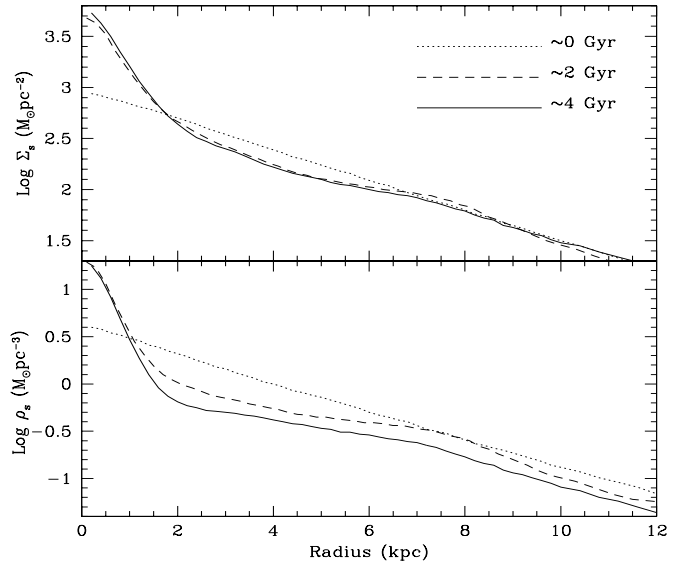}
\caption{Evolution of the azimuthally averaged disc surface density
(top) and the central volume density (bottom) for the model
$C$. The epochs are shown in the top panel.  The volume density and
dispersion velocities at each radius were calculated inside slabs of
height equal to 0.8h$_z$(R).}
\end{figure}

\section[]{EVOLUTION OF THE  COARSELY GRAINED PHASE-SPACE DENSITY }

Figure 3 shows the evolution of the surface and the volume
densities averaged azimuthally for the simulation of model $C$,
as an example. Dotted lines correspond to the 
beginning of the simulation, and the short dashed and solid lines 
are for moments of time separated approximately by 2~Gyr. The effect 
of bar evolution on the disc surface density  is significant: matter 
is accumulated at the centre while the slope of the outer disc becomes 
shallower than the initial slope, with the disc scale-length increasing 
by $\sim 30\%$ (VK03). The volume density shows similar behavior,
but including also the disc vertical expansion with time. 

  The disc gets hotter and thicker as clearly demonstrated in
 Fig. 4, which shows the evolution of the three components of the
 velocity dispersion as well as h$_z$(R) for the same model $C$. 
 The disc heating is very
 large in the central $2-4$~kpc region where the bar forms. The heating 
 happens also in the outer disc but to a much lesser degree. Here the heating 
 is due to spiral waves, which form in the initially unstable disc. The waves
 gradually decay and heat the disc, but most of the heating occurs in
 the plane of the disc: the radial and tangential rms velocities
 increase substantially more than the vertical rms velocity, which
 changes by 25-30 percent over 4~Gyrs (see also VK03).

Finally, the evolution of the ``observational'' radial phase-space density
profile, \fp, of model $C$ is shown in Fig. 5(a) . Because the initial disc 
of this model has a constant $Q$, the initial \fp(R) profile is
$U-$shaped: the maximum at the centre is followed by a minimum
at the inner disk. The steepness of the central maximum depends on the 
behavior of $\kappa$(R) at small radii. For model $C$, the inner 
\fp(R) profile is almost flat. In Fig. 5 (panels b, c and d) are shown
also the evolution of \fp(R)~ for models $D_{\rm cs}$, $D_{\rm hs}$, and $A_1$,
respectively. The initial inner \fp\ profile of the first two models are 
significantly steeper than for model $C$, while for model $A_1$, 
\fp~ decreases toward the centre. In the latter case, 
$\sigma_R^2(R)\propto \exp(-R/$h$_d)$ was assumed initially 
instead of $Q=const$ (see \S 1.1).

\begin{figure}
\vspace{10.cm}
\includegraphics{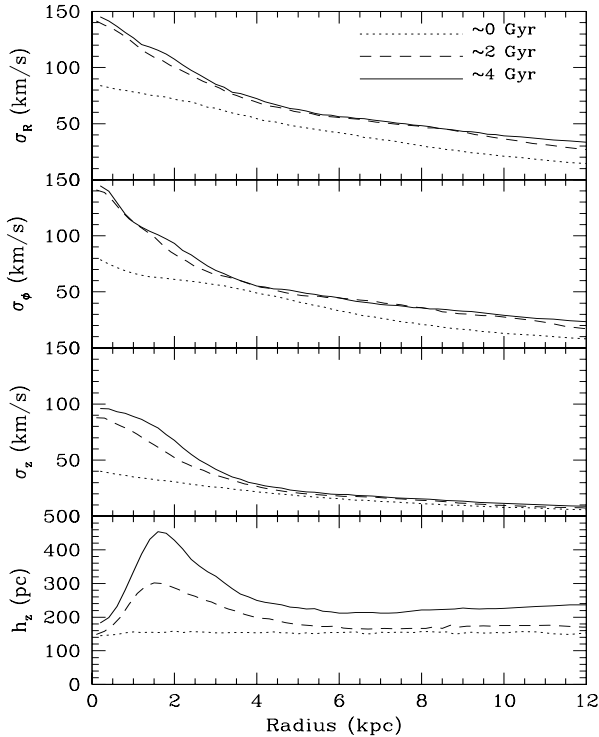}
\caption{{\it Upper panels:} Evolution of the radial, tangential and vertical 
velocity-dispersion radial profiles for model $C$. {\it Bottom panel:} Evolution 
of the scale height h$_z$.}
\end{figure}

One clearly sees that the macroscopic (observational) azimuthally averaged
phase-space density decreases with time along the whole disc of all 
the models. In the outer parts of the disc the bar and the spiral
arms heat and thicken the disc. Here the surface density at each
radius remains almost constant. The disc heating explains why
\fp(R) decreases at all radii.  In the inner disc, the \fp~ profile
of all the models develops a valley whose depth increases with time. For
example, for model $C$ the minimum of this valley is at $\sim 1.5-2$kpc 
and this is close to the radius where the central mass concentration ends. 
Around this radius one observes the maximum radial mass exchange as well 
as the maximum vertical heating and disc thickening due to the bar. 
Here the outer region of the bar produces a large phase-space
mixing -- larger than in the centre. As the result, \fp\ at some radius
inside the central mass concentration is higher than \fp\ measured at the 
inner disc. The largest changes in the \fp(R) profile, as well as in other
quantities (such as $h_z$ and rms velocities) occur during typically the 
first $\sim 1$ Gyr of evolution. The evolution continues, but much slower at
later moments.

\section{Bulge-like structure formation and robustness of the results}

The bar-driven evolution produces a dense central concentration (or
even a pseudobulge as is the case of model $D_{\rm cs}$) with a slope
steeper than the original one. 
Figure 3 clearly illustrates this point for model $C$. For almost all of 
our simulations the surface density in the inner $\sim 2$kpc region is 
well approximated by a S\'ersic profile (S\'ersic 1968) with slope 
index $n<4$. For example, for model $C$, $n\approx 1$.  This is similar 
to what is observed for bars and bulges of late type galaxies (e.g., de Jong
1996; Graham 2001; Mac Arthur et al. 2003; Hunt et al. 2004).  

Comparison of different models indicate remarkable similarities in the
evolution and the shape of the phase-space density profile \fp(R). 
In all the models \fp(R) has a maximum at the centre followed by a deep 
minimum at $\sim 1-2$kpc, where the corresponding outer bars live (see Fig. 5). 
It seems that the results do not depend on the numerical resolution and are 
not particularly sensitive to initial conditions.  For example, for model 
$A_1$ the resolution is lower than for model $C$, and $Q$ was not assumed 
constant initially.  Still, the evolution and shape of the
\fp\ profile are similar to those of the model $C$ (panels a and d in Fig.
5). Although resolution is an important factor in simulations aimed to
explore gravitational instabilities of {\it thin} discs embedded in
large hot haloes (O'Neill \& Dubinsky 2003; VK03), we find that the
shape and evolution of \fp(R)\ is qualitatively the same in
simulations with different resolutions. 

When we look closely at the results, we clearly see differences
between models.  For example, the minimum of \fp(R)~ is at different
radii and the depth of the minimum varies from model to model. The
smallest radius and the deepest minimum are attained by model $D_{\rm cs}$,
where the bar is dissolved and a pseudobulge forms. The differences
in the evolution of \fp(R) are expected because the lengths and strengths 
of bars are different in different models. Yet, it seems that the overall 
generic shape of \fp~ after evolution is a robust prediction of the secular 
collisionless scenario.

\begin{figure}
\vspace{13cm}
\includegraphics{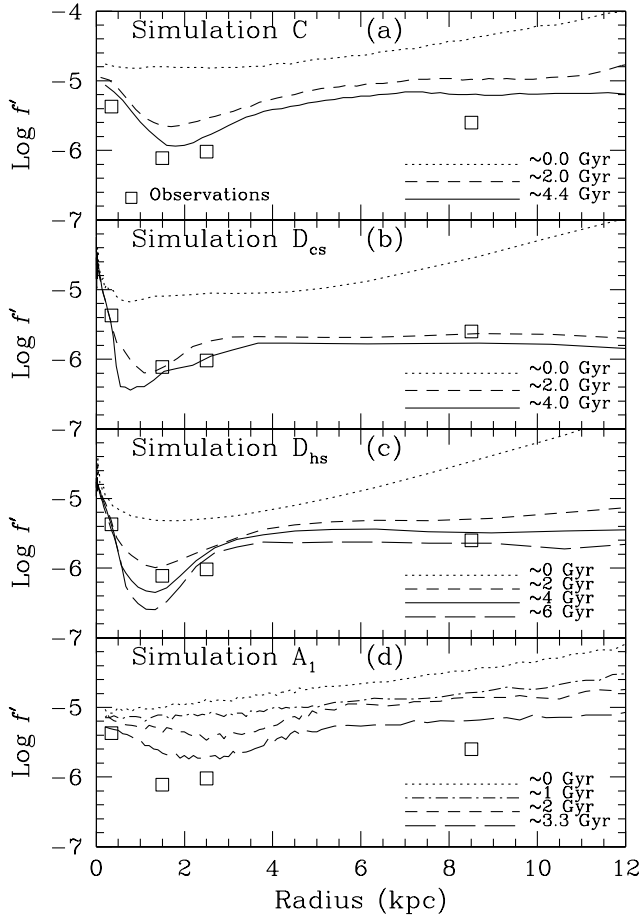}
\caption{{\it (a)} Evolution of the azimuthally averaged radial \fp\ 
($=\rho/\sigma_R \sigma_{\phi} \sigma_z$ \fdunit) profile for model 
$C$ simulation. 
Open squares are estimates corresponding to our Galaxy.
{\it (b)} Same as (a) but for the simulation of model $D_{\rm cs}$, where
the bar dissolves and a bulge forms. {\it (c)} Same as (a) but
for the high-resolution simulation of model $D_{\rm hs}$, where
the bar weakened but it did not dissolve. Note that for this simulation 
\fp(R) is shown until 6 Gyr. {\it (d)} Same as (a) but for simulation 
of model $A_1$, where h$_z$=const instead of $Q$=const at the beginning.
For this simulation \fp(R) is shown only until 3.3 Gyr. The evolution of 
\fp(R) is qualitatively similar in all the simulations, attaining a 
minimum value in the inner disk. }

\end{figure} 

\section{Comparison with observations}


Wyse (1998) presented estimates of \fb\ and \fd\ for the Galaxy 
from available observational information. We use more recent 
data to give updated estimates. We measure \fb\ and \fd\ 
at the typical radii r$_e$/2 (r$_e$ is the effective bulge radius, 
r$_e\approx 0.7$ kpc, see Tremaine et al. 2002) and \Rdi=2.5 kpc, 
respectively. The bulge and inner disc stellar volume densities are 
estimated from the galaxy model used in Bissantz \& Gerhard (2002), 
where the parameters were fixed from fittings to the dust-corrected 
COBE-DIRBE $L-$band maps (Spergel et al. 1995). Taking into account 
the bulge ellipticity and averaging vertically the disc density within 
h$_z=330$ pc (inner disc scale height, Chen et al. 2001), we obtain 
$\rho_b($r$_e/2)=6.7 \msunpc$ and $\rho(\Rdi)= 0.31\msunpc$. 
Regarding velocity dispersions, for the bulge we use an interpolation 
of measured $\sigma_{p}$'s at different projected galactocentric 
distances as compiled by Tremaine et al. (2002) ($\sigma_{p}$(r$_e$/2)
= 116.7 km/s, corrected for ellipticity),  and we assume velocity 
isotropy and that $\sigma_r\approx \sigma_{p}$. This approximation is 
valid for S\'ersic profiles within $0.1\lesssim r/r_e \lesssim 10$ 
(Ciotti 1991). 

For the disc, the three velocity dispersions at \Rdi\ are calculated 
by using the radial profiles given in Lewis \& Freeman (1989).
Thus, at \Rdi, $\sigma_R=78.9$ km/s, $\sigma_{\phi}= 73.1$ km/s, and 
$\sigma_z=41.8$ ($\sigma_z=0.53\sigma_R$ was assumed). For completeness, 
we also calculate \fp\ in the outer bulge (at 1.5 kpc) and in the solar 
neighborhood (8.5 kpc). For the former, the same Bissantz \& Gerhard 
(2002) bulge model and the Tremaine et al. (2002) recompilation for 
$\sigma_{p}$ were used. For the latter, we use the local estimate for the 
stellar density, $\rho_s=0.09$ \msunpc\ (Holmberg \& Flynn 2000), and the 
velocity dispersion profiles from Lewis \& Friedmann (1989). 

The values of $\fb (r_e/2)$, $\fb(1.5$kpc$)$, \fd\ and $f'_{\odot}$ 
($4.3\ 10^{-6}$, $7.8\ 10^{-7}$, $9.6\ 10^{-7}$, and $2.8\ 10^{-6}$ \fdunit, 
respectively) are indicated with empty squares in Fig. 5. 
The qualitative agreement in the shape of \fp~ between observations and 
numerical predictions is remarkable, in spite of the fact that the 
observations have large uncertainties and the models do not include gas, 
SF processes or gas infall and minor mergers. Note also that  \fp(R) 
tends to stabilize after $1-2$ Gyr, although one sees still changes after 
this period. We consider that the collisionless secular scenario (not including
dissipative physics) is a good approximation for Milky Way-like galaxies
for their last 4-7 Gyr. Before this time, the discs were much more
gaseous and dissipative phenomena should be taken into account. 

The models predict that the phase-space density decreases with time
even at large radii. This decline of \fp~ in outer regions is produced 
by spiral waves, which develop in the unstable disc.  In real galaxies the 
same effect
should be produced by real spiral waves and possibly by molecular
clouds. It is interesting to compare the models with what is observed
for our Galaxy at the solar neighborhood. At $R\sim 8$ kpc in the
models \fp~ roughly decreases by a factor 5-10 during 4-7~Gyrs of
evolution. Studies in the solar neighborhood show that the components
of the stellar velocity dispersion increase with the age of
stars. This is the so called age-velocity relation (for recent
estimates see Rocha-Pinto et al. 2004 and the references
therein). Observations indicate that each of the three velocity 
dispersion components increases by $20-60\%$ during the 
last $\sim 6-7$Gyr of galactic evolution. If this
is an indication how the whole stellar population evolves, we can make
rough estimates of the evolution of \fp~. For an equilibrium disc with
stellar surface density $\Sigma$ and total rms velocity $\sigma$, the
phase-space density scales as $\fp\propto \Sigma^2/\sigma^5$. If for
the sake of argument we assume that the rms velocity of the whole
stellar population increased by factor of 2 during the last 6~Gyrs
without substantial change in the stellar surface density, then we expect
decline in \fp~ by 32 times. Likely increase in the stellar surface 
density $\Sigma$ (e.g., Hern\'andez, Avila-Reese \& Firmani 2001)
should reduce this very large factor. Yet, naively one does expect for 
our Galaxy a large drop in \fp~ just as our models predict.

Finding \fp\ for external galaxies is not easy, specially when it 
concerns the velocity dispersion. Here we  estimate the 
ratios of \fb, calculated at r$_e$/2, to \fd, calculated at 0.8h$_d$ 
for four spiral galaxies of different morphological types (Sa - Sbc) 
studied by Shapiro et al. (2003). We 
use the $K-$ and $I-$band surface brightness profiles and the $\sigma_{p}$ 
profiles reported in Shapiro et al. (2004). The disc volume density is 
assumed to be proportional to the surface brightness divided by 2h$_z$ 
(h$_z$ is set equal to 0.125h$_d$, Kregel 2003). We also use the disc 
($\sigma_R,\sigma_{\phi},\sigma_z $) profiles, which Shapiro et al. 
fit to their spectroscopic observations. The bulge luminosity density profile 
is  estimated as follows: (i) the surface brightness profiles reported in 
Shapiro et al. are decomposed in a S\'ersic bulge and an exponential
disk, (ii) spherical symmetry is assumed for the bulge, (iii)
its luminosity density profile is calculated with the found S\'ersic 
parameters by using the approximations given in Lima-Neto, Gerbal \& Marquez 
(1999). We use $\sigma_p^3$ for the bulge, 
assuming an isotropic velocity-dispersion tensor and that $\sigma_r = 
\sigma_p$; therefore, our $f_b$ is a lower limit. To pass from 
luminosity to mass, we assume that the M/L ratios in the $I$ and $K$ band 
are 2 and 1.5 times larger for the bulge than for the disc, respectively.

Table 2 shows the estimates of \fb(r$_e$/2)/\fd(0.8h$_d$) for the spirals
from Shapiro et al. (2004). The \fb(r$_e$/2)/\fd(0.8h$_d$) ratio is
indeed $>1$ for three galaxies and close to one for the last one. NGC
4030 has the latest type (Sbc) among the four and probably it was not
affected significantly by the secular evolution.  Overall, the results
are consistent with what we find for the Galaxy. 

\begin{table}
  \caption{Bulge S\'ersic index n and phase-space density bulge-to-disc ratio
for four galaxies.}
  \begin{tabular}{@{}lccc}
  \hline
   Name & Type & n & \fb(r$_e$/2)/\fd(0.8h$_d$) \\
 \hline
 NGC 1068   & Sb    & 2.1 & $>6.7$  \\
 NGC 2460   & Sa    & 1.7 & $>3.2$\\
 NGC 2775   & Sab   & 1.8 & $>1.3$\\
 NGC 4030   & Sbc   & 2.0 & $>0.9$ \\
\hline
\end{tabular}

\end{table}
 
\section{Conclusions}

We studied the evolution of the observational measure of the coarsely
grained phase-space density, \fp, in high-resolution N-body simulations 
of Galaxy-like models embedded in live CDM halos. In our models the 
initially thin stellar disc is unstable.  
As the system evolves, a bar with almost exponential density profile is
produced. The bar redistributes matter in such a way that the disc ends with
a high accumulation of mass in the centre and and extended outer disc with 
a density profile shallower than the exponential law. During the secular 
evolution, the disc is also dynamically heated and thickened, mainly in the 
inner parts where a bulge-like structure (peanut-shaped bar or pseudobulge) 
arises.

The secular evolution produces dramatic changes in the radial distribution
of the coarsely grained phase-space density \fp(R). As the
disc is heated and expanded vertically, \fp(R) decreases at every radius $R$.  
The outer region of the bar produces a large phase mixing in the inner
disc  --larger than at the centre. As the result, the \fp(R) profile 
develops an increasing with time valley, with a pronounced minimum in 
the inner disc, where the central, bulge-like mass concentration ends. 
In this region the vertical heating and the radial mass exchange in the 
disc are maximum.
Our results on the evolution and shape of the \fp\ profile
are qualitatively robust against initial conditions and assumptions,
numerical resolution, and the way of measuring the volume
density and dispersion velocities.

We conclude that the secular evolution of a {\it collisionless}
galactic disc is able to form a thick, dynamically hot, central mass
concentration (eventually a pseudobulge), where the phase-space density 
is much higher than in the inner disc.  Using observational data we have 
estimated \fp~ at several radii for the Galaxy. In particular we estimated 
the \fp~ bulge-to-inner disc ratio.  The qualitative agreement with our
numerical results is remarkable. Therefore, the
secular evolution of a collisionless disc yields a radial
coarsely grained phase-space density profile in agreement with that it
is observationally inferred for the Galaxy. 
The phase-space density constraints favor the bulge secular
formation scenario. The inclusion of other important physical
ingredients, as gas dissipative effects and satellite accretion, will 
likely enhance the secular evolution of disc models.


\section*{Acknowledgments}
We are grateful to J. Gersen for sending to us their observational
data in electronic form, and to S. Courteau for useful discussions.
We acknowledge J. Wadsley and T. Quinn for allowing us to mention
results from a simulation carried out with the the Gasoline code 
before its publication. We thank the referee for useful comments.
This work was partially funded by CONACyT grant 40096-F, and by a 
DEGEP UNAM grant to A.~C. He also thanks CONACyT for a graduate fellowship. 
In addition, O. V. acknowledges support by NSF ITR grant NSF-0205413, 
and A. K. acknowledges support by NASA and NSF grants.

\bsp

\label{lastpage}

\end{document}